\documentclass{JINST}
\usepackage{overpic}

\title{Luminosity Monitoring in ATLAS \\ with MPX Detectors}

\author{A.~Sopczak$^a$ on behalf of the ATLAS and Medipix-2 Collaborations\\
\llap{$^a$} Institute of Experimental and Applied Physics, \\
Czech Technical University in Prague, \\ 
Czech Republic\\
E-mail: \email{andre.sopczak@cern.ch}}

\abstract{The ATLAS-MPX detectors are based on  the 
Medipix2 silicon devices designed by CERN for the detection of 
multiple types of radiation. 
Sixteen such detectors were successfully operated in the ATLAS 
detector at the LHC and collected data independently of the ATLAS 
data-recording chain from 2008 to 2013.
Each  ATLAS-MPX detector provides  separate  measurements of 
the bunch-integrated LHC luminosity. 
An internal consistency for luminosity monitoring of about 2\% was
demonstrated.
In addition, the MPX devices close to the beam are sensitive enough 
to provide relative-luminosity measurements during van der Meer 
calibration scans, in a low-luminosity regime that lies below the 
sensitivity of the ATLAS calorimeter-based bunch-integrating luminometers.
Preliminary results from these luminosity studies are presented 
for 2012 data taken at $\sqrt{s}=8$~TeV proton-proton collisions.}


\begin{document}

\section{Introduction}
\label{sec:introduction}

Pixel detectors have a widespread use in various research 
fields and applications such  as  particle and nuclear physics, as well as
medical and industrial imaging. A Medipix2 (MPX) 
chip developed at CERN in the framework of the Medipix2 
Collaboration~\cite{medipixCollaboration:2013} is an example of such a device. The data  of 
the MPX devices is  recorded  in  frames  which  contain the 
status of all the  65,536 pixels  after a given exposure time  of 
the order of~1\,s to~100\,s. The dead time after each frame is about 6~s 
for data readout.  The  MPX  devices can effectively be 
used for continuous measurements of the composition of 
complex radiation fields~\cite{mpx2ProjectProposal:2006}. A network of  MPX devices was 
installed within the ATLAS experiment~\cite{atlasCollaboration:2013} at CERN.  
It was  primarily designed to provide real-time measurements of the composition 
(photons, neutrons and charged particles) and of the spectral characteristics of 
the radiation environment inside the ATLAS detector. 
The  large dynamic range of the  MPX  devices allows  the measurement 
of the natural radiation background,  and of the induced activity  during 
and after  collision periods. The devices record the  decay of 
radioactive nuclei generated in  ATLAS  during LHC 
collisions.    This measurement  of the LHC-generated
radiation  field composition  permits the validation of  radiation 
simulation studies. Recently,  descriptions and  results from 
2008-2011  MPX  radiation field measurements  have been 
released~\cite{analysisRadiaField:2013}.

The use of the  MPX  network for relative-luminosity 
measurements in proton-proton collisions has been studied in 
detail  and the results are summarized in this report. 
Sixteen devices placed in the \mbox{ATLAS} detector are read out independently.
Out of these, two had high noise and one was located in the data-acquisition room, 
so only thirteen are used in this analysis.
Table~\ref{tab:mpx_detectors}  lists  the  location  and number of registered events 
(clusters) per sensor area and per unit integrated luminosity  of these thirteen  MPX 
devices. 

\begin{table}[hbtp]
\caption{MPX device locations with respect to the interaction point. 
Z is the longitudinal distance from the interaction point and
R is the distance from the beam axis. 
The number of measured MPX clusters per sensor area 
and per unit integrated luminosity is given.
Only devices with low cluster rates are used for the neutron counting analysis as indicated. 
The devices are grouped according to their ordering 
in Fig.~3
as indicated by the horizontal lines.}
	\label{tab:mpx_detectors}
\vspace*{-2mm}
\begin{center}
\renewcommand{\arraystretch}{1.3}
             \begin{tabular}{crrccc}
                        \hline\hline
Device & Z~ & R~  &Measured MPX clusters & Used for \\
                & [m]   & [m]    &per sensor area and per unit  & neutron\\
                         & & & luminosity $[\rm cm^{-2}/ nb^{-1}]$ & counting\\
                        \hline
                        MPX01 & 3.42 & 0.77  & 55000   & No\\
                        MPX13 & $-3.42$ & 2.44 & 380  & No\\
                        MPX02 & 3.42 & 2.50  & 230     & No\\
                        MPX03 & 2.94 & 3.57 & 31      & No\\
                        MPX06 & 7.20 & 3.36 & 20      & Yes\\   
                        \hline
                        MPX05 & 7.20 & 2.36 & 47    & No\\
                        MPX08 & 4.02 & 4.40 & 1.2   & Yes\\
                        MPX07 & 0.35 & 4.59 & 0.45  & Yes\\
                        \hline
                        MPX04 & 7.12 & 1.30  & 110   & No\\
                        MPX09 & 15.39 & 1.56 & 5.8   & Yes\\
                        MPX12 & 7.23 & 6.25  & 3.9 & Yes\\
                        MPX10 & 22.88 & 5.19 & 1.0 & Yes\\
                        MPX11 & 4.86 & 16.69 & 0.30  & Yes\\
                        \hline\hline
                \end{tabular}
\end{center}
\end{table}

\begin{figure}
\begin{overpic}[width=0.49\linewidth]{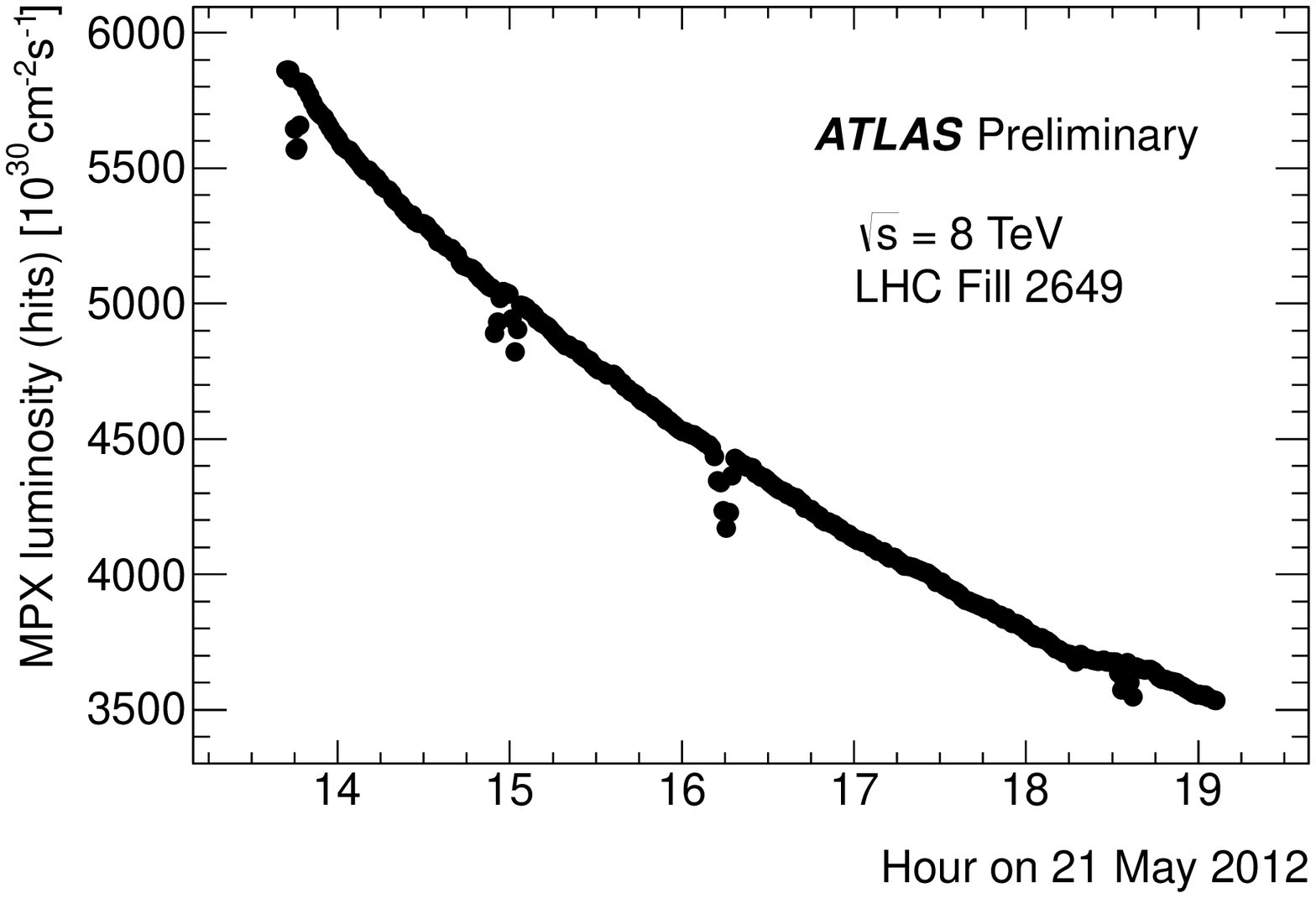}
	\put(86, 59.0){(a)}
\end{overpic}
\begin{overpic}[width=0.49\linewidth]{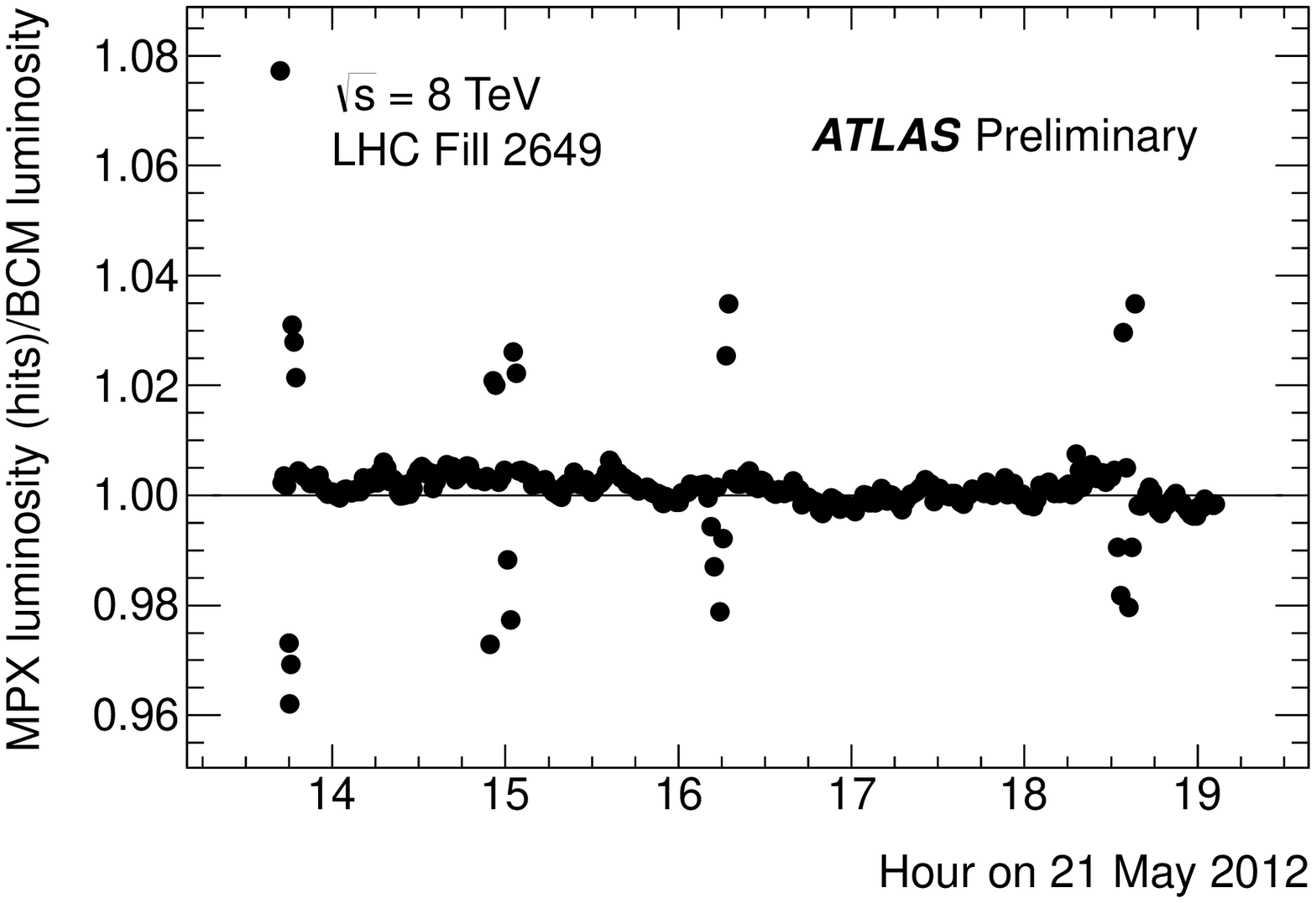}
	\put(86.0, 59.0){(b)}
\end{overpic}
\vspace*{-3mm}
\caption{Time history of (a) the MPX luminosity and (b) the MPX/BCM luminosity ratio.
The small luminosity dips apparent in (a), 
and the associated outliers in (b), 
correspond to times when the LHC operators perform small-amplitude beam-separation scans 
to optimize the luminosity. During these brief intervals, the luminosity varies more rapidly 
than can be sampled by the online luminosity and MPX data-taking infrastructure, 
resulting in unphysical variations of the apparent MPX/BCM luminosity ratio.}
\label{fig:mpxMayRun}
\end{figure}

\begin{figure}
\vspace*{-2mm}
\centering
\includegraphics[width=0.49\linewidth]{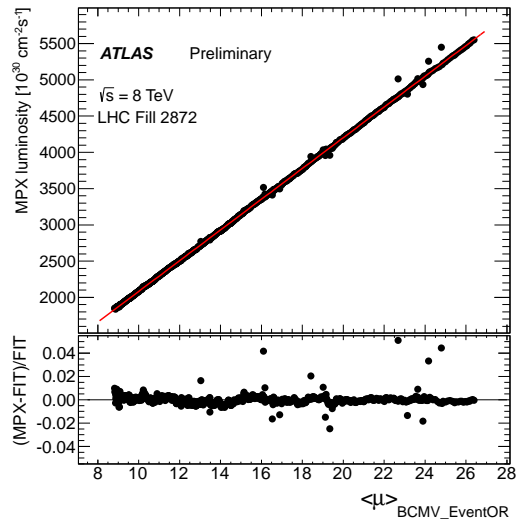}
\vspace*{-3mm}
\caption{MPX01 hit luminosity vs. bunch-averaged number $\langle \mu \rangle$ of inelastic interactions 
per crossing reported by the reference ATLAS luminosity algorithm ($\rm BCMV\_EventOR$). 
This LHC fill was chosen because the data span a particularly wide range of 
$\langle \mu\rangle $ values.
The residual of a linear fit to the data is shown in the bottom frame. 
The outliers correspond to times when the LHC operators performed small amplitude 
beam-separation scans to optimize collisions, during which the luminosity varies more 
rapidly than can be reliably sampled by the online luminosity and MPX data-taking infrastructure.}
\label{fig:mpx01HitLumi}
\end{figure}

The luminosity monitoring is based on the measurement of 
the  integrated  rate of particles  interacting  with  the  300 $\mu$m thick
silicon-sensor (hit rate  measurement in the so-called counting 
mode~\cite{analysisRadiaField:2013}).  The hit rates per frame are converted into 
luminosity  using a normalization factor  for each MPX device. 
The integrated hit rate corresponds to the integrated luminosity reported 
by the reference ATLAS luminosity algorithm ($\rm BCMV\_EventOR$ 
algorithm~\cite{improvedLumiDet:2013} using the 
Beam Conditions Monitor, BCM~\cite{BCM2008}),
during an ATLAS run chosen as a reference (LHC fill 2649, 21 May 2012).
The MPX devices operated reliably starting from 2008 even  in places 
where the radiation level was high. The advantage of the devices is 
that they have 65,536  independent channels (pixels). 
Figure~\ref{fig:mpxMayRun} shows an
example of the hit luminosity measured with MPX01 and the 
ratio of the MPX luminosity with respect to the ATLAS 
BCM reference  luminosity. 
The relative linearity of the MPX and BCM luminosity measurements 
is illustrated by Fig.~\ref{fig:mpx01HitLumi}.
In a run selected for covering a particularly wide span of bunch-luminosity values, 
the residual of a linear fit displays no systematic non-linearities down to the 
sub-percent level, for a bunch-averaged pile-up parameter $\langle \mu \rangle$ varying 
from 8 to 26 inelastic interactions per bunch crossing.
The purpose of the plot is to show that there
is a linear relationship between the BCM and MPX luminosities.
Since the luminosity range is limited, it is
possible that the linearity is not perfect
when extrapolating to zero luminosity.

In addition to the counting mode,  devices with 
low cluster rates
(MPX06 to MPX12) can also be operated in the so-called  
tracking mode~\cite{analysisRadiaField:2013},
 which allows recognition and 
categorization of different types of radiation by their 
characteristic track signature (cluster shape).  
This mode is used for luminosity measurements by counting thermal neutrons.

In the process of performing the luminosity determination with MPX 
devices, the activation of the ATLAS detector material has 
been investigated and found to have a negligible effect. 

\section{MPX Luminosity from Hit Counting}
\label{sec:lumi_hit_counting}

An  advantage of the MPX network for luminosity 
measurements is that there are multiple independent devices at 
different locations installed in the ATLAS detector  and  in the 
cavern.  This allows comparative studies to cross-check independent  hit  luminosity measurements by the 
thirteen MPX devices which have been used for this analysis.

In order to  ensure that the results do not depend on  the 
analysis method used,  three independent analyses have been 
performed and the results  are  compared.  
The MPX luminosity data are saved in a different time window (acquisition frame time) 
than the ATLAS luminosity. The basic time unit for storing ATLAS luminosity for 
physics use is the Luminosity Block (LB), which is typically one minute long. 
The three methods differ in the procedure and criteria for noisy pixel removal 
(a potential source of systematic uncertainty in hit-counting mode), 
and for scaling the raw hit counts from the exposure window of each MPX device 
to the duration of the matching ATLAS LB. The three methods are described as follows.
\begin{itemize}
\item{Method 1: A noisy pixel is defined as having a counting rate that differs
from the mean rate by more than five standard deviations.
If a pixel is found to be noisy in a 24 hour reference period 
(chosen as the day in 2012 that had the largest number of noisy pixels) 
it is removed from the entire 2012 data-taking period.
A linear interpolation is made between the rates in different frames. The value at
the middle of a LB obtained from this interpolation is used.
}
\item{Method 2: Noisy pixel removal  is done  frame by frame,
i.e. in each frame a different set of noisy pixels is removed.
Noisy pixels are defined by having a count rate that differs from the mean by more
than a luminosity-dependent threshold.
The MPX luminosity from frames falling within a LB is used without an
interpolation. 
A correction is made for the relative duration of the MPX frames and of the LB.
}
\item{Method 3: Noisy pixel removal  is done  frame by frame.
The counts of 15 frames (the frame under investigation and 7 frames before and after) 
are summed and a pixel is removed if the sum of these counts is above a threshold.
An interpolation of the frame hit rate at the time of each LB is 
done as in method~1.}
\end{itemize}
In MPX01, all three methods show a significant increase of 
the number of noisy pixels with time. As this sensor sustains the largest particle 
flux of the analysed devices, this observation could be an indication of radiation damage. 
This might result from defects in either the silicon or the readout chip. 
Using method~1 (method~2), the number of noisy pixels in MPX01 increases from less 
than 10 (300) in April 2012 to about 300 (1800) at the end of November 2012.

The luminosities determined by the three methods were compared in
short (frame-by-frame) and long (7 months) time periods. 
Depending on the MPX device considered, the frame-by-frame agreement 
varies from a few percent to less than 0.1\% (for MPX01). 
For the long-term comparison, the MPX hit luminosities from the three 
methods were studied with respect to the BCM reference luminosity. 
About the same variations of the MPX/BCM luminosity ratio 
as a function of time were observed with the different methods.

\begin{figure}[!t]
\begin{overpic}[width=0.49\linewidth]{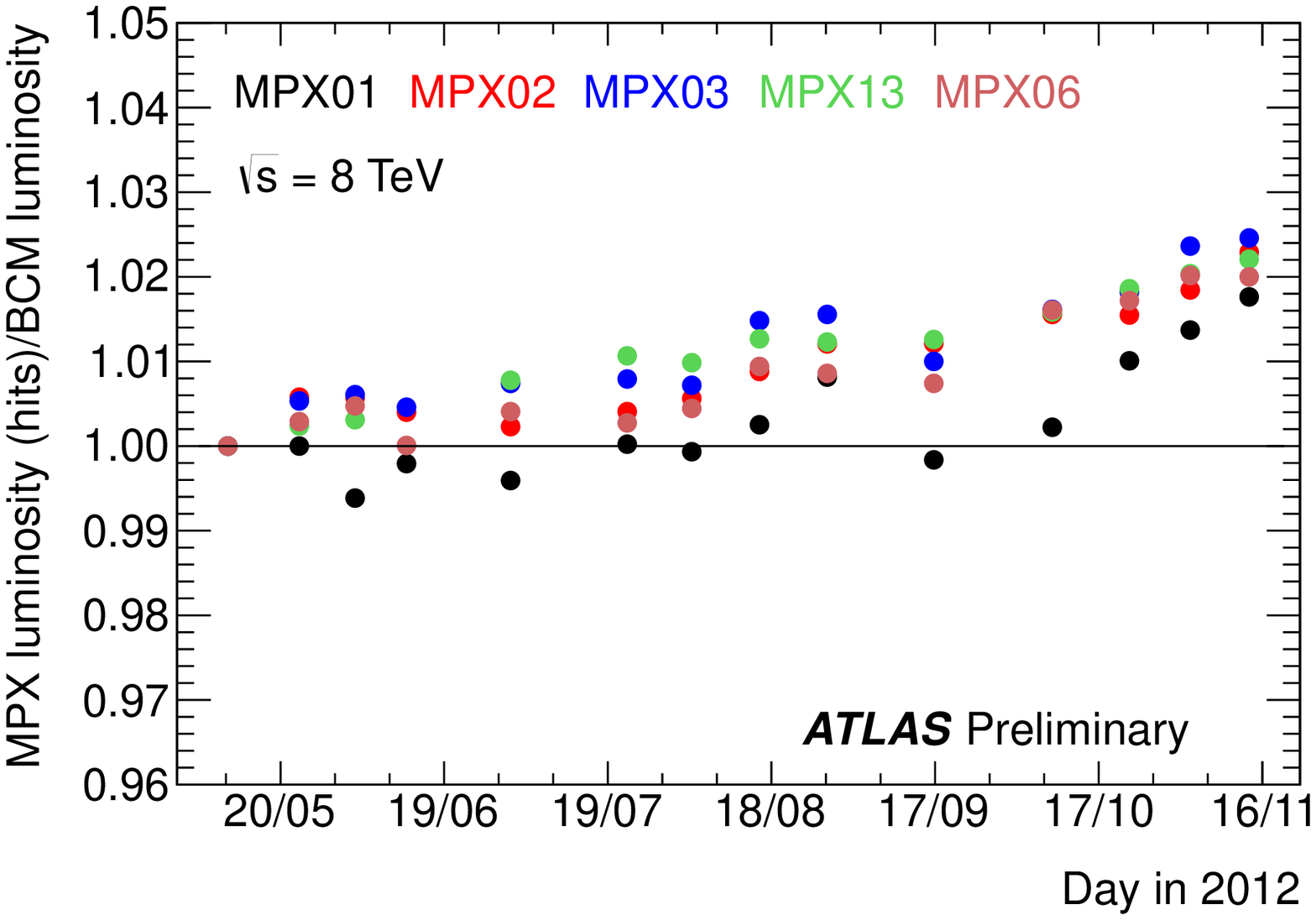}
	\put(86, 59.0){(a)}
\end{overpic}
\begin{overpic}[width=0.49\linewidth]{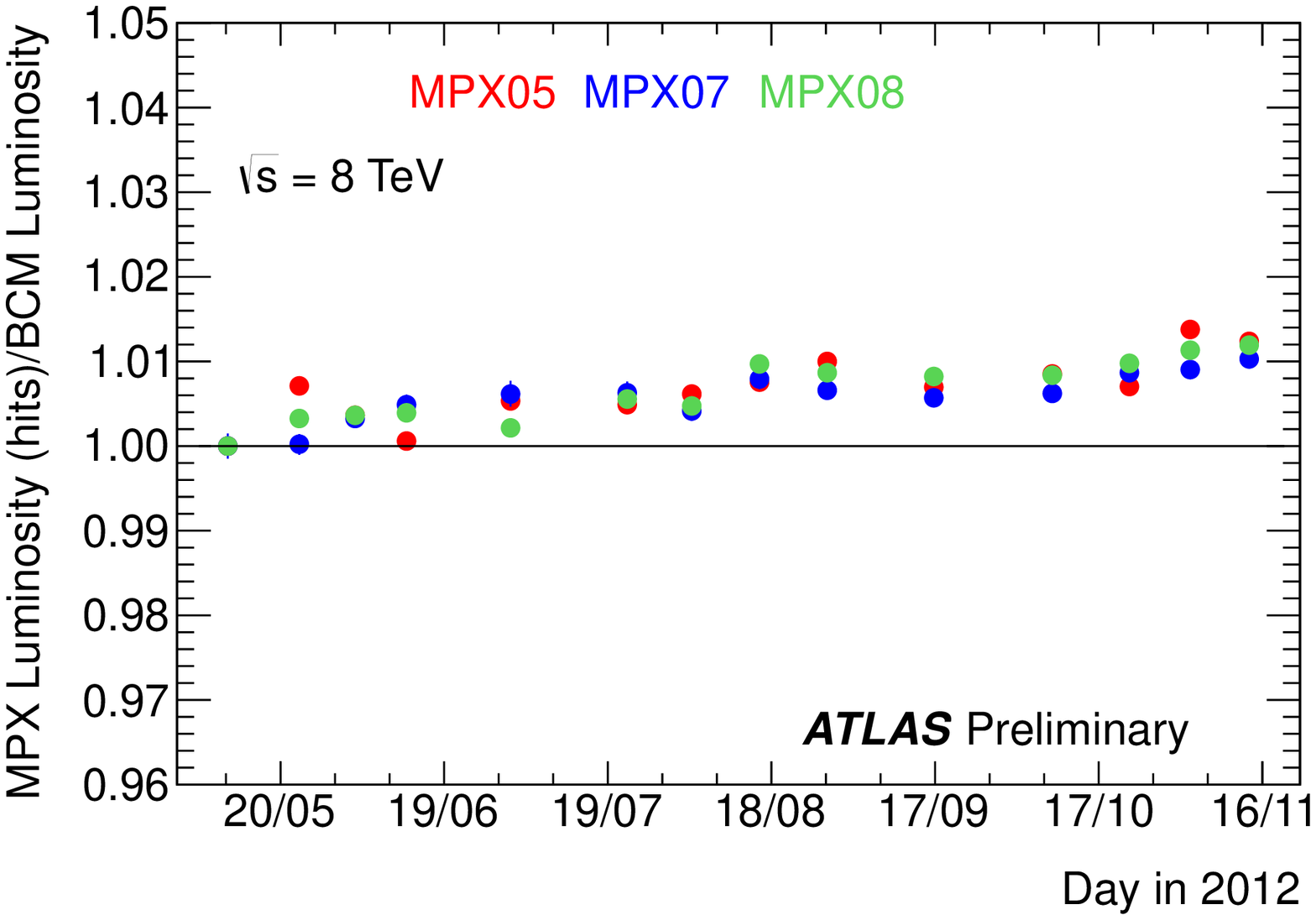}
	\put(86, 59.0){(b)}
\end{overpic}
\begin{center}
\begin{overpic}[width=0.49\linewidth]{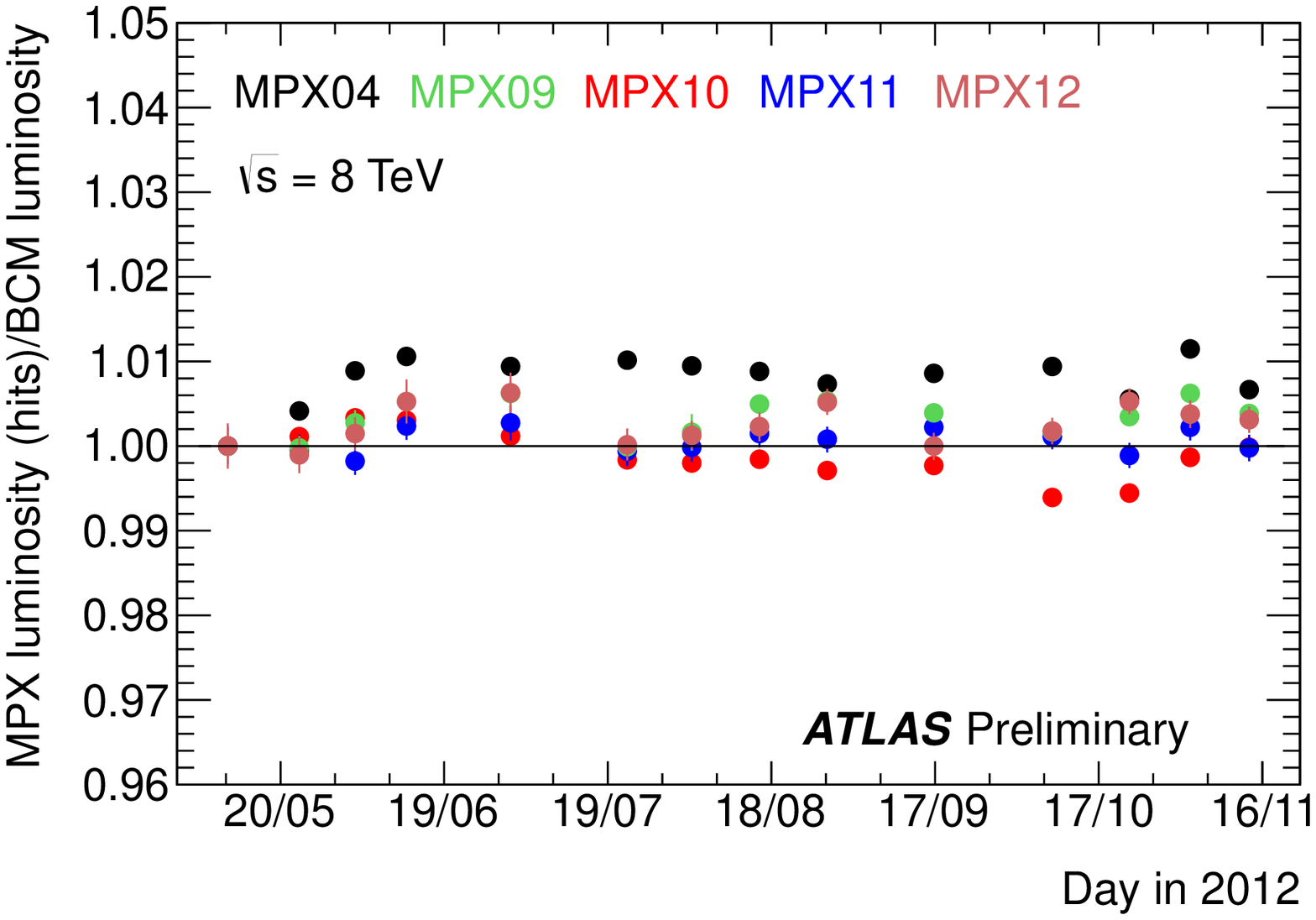}
	\put(86, 59.0){(c)}
\end{overpic}
\end{center}
\vspace*{-5mm}
\caption{Fractional deviation of the luminosity reported by different MPX devices 
from the ATLAS reference luminosity (BCM) as a function of time in 2012. 
The ratio is shown for 14 time periods. 
For each device the time periods are chosen to have the same number of frames. 
The statistical uncertainties are comparable to or smaller than the size 
of the markers.
The data are scaled such that the value of the first bin is unity.}
\label{fig:summaryHits}
\end{figure}

In order to study the relative long-term stability of the luminosity 
measurements, 
the ratio of the  MPX  luminosity 
to the ATLAS reference luminosity, from May to November 2012, 
is shown in Fig.~\ref{fig:summaryHits} 
with a binning in 14 time periods. 
For this analysis we have taken method~1 as a baseline for hit counting.

The ratios measured by different MPX devices are divided in three groups.
The devices shown in Fig.~\ref{fig:summaryHits}a display an increase of about 2\% over the 2012 
running period, while the ratios in Figs.~\ref{fig:summaryHits}b and~\ref{fig:summaryHits}c 
show an increase of 1\% or less over the same period. 
The three groups of devices are listed in Table~\ref{tab:mpx_detectors} together with their
cluster
rates and although there is some indication that high rate 
devices show an increase while low rate devices do not, 
the picture is not conclusive.
The slightly different time dependence of result from different MPX 
devices is not understood 
at present and  the conclusion is therefore that 
the long-term internal consistency of the MPX system is at the 2\% level.

\section{MPX Luminosity from Thermal Neutron Counting }
\label{sec:lumi_th_neutron_counting}

Thermal neutrons are detected by MPX devices 
via $\mathrm{{}^{6}Li(n,\alpha){}^{3}H}$ reactions 
in a $\mathrm{{}^{6}LiF}$ converter layer~\cite{analysisRadiaField:2013}. 
In the MPX tracking mode tritons and alpha 
particles are registered as so-called heavy blobs (large round-shaped pixel 
clusters). The number of heavy blobs below a $\mathrm{{}^{6}LiF}$ converter 
is used to calculate the thermal neutron fluence using a detection 
efficiency obtained for each MPX device in a standardized 
isotropic thermal neutron field. 
The typical detection efficiency of thermal 
neutrons is 1\%~\cite{analysisRadiaField:2013}.  

The chosen acquisition times allow the measurement of the thermal neutron counts 
for MPX06 to MPX12 as for these devices the total cluster occupancy per frame 
has a correctable overlap, i.e.,  at most  a few  hundred  clusters  per frame.
A dedicated study was performed to determine the misidentification of heavy 
blobs which are lost due to the overlap with other 
clusters~\cite{analysisRadiaField:2013}. 
The resulting correction factors, which are specific to each MPX device, depend on
the number of clusters per frame (i.e. on the luminosity and on the device location); 
they vary between one and about two. 
The precision of these correction factors was estimated to be below 1\% 
using different data sets.
These results were also confirmed by an independent analytic study, 
that calculated the cluster 
overlap probability for a simple cluster type and found very good agreement with the 
experimentally determined overlap probability.

\begin{figure}[tp]
\centering
\includegraphics[width=0.49\linewidth]{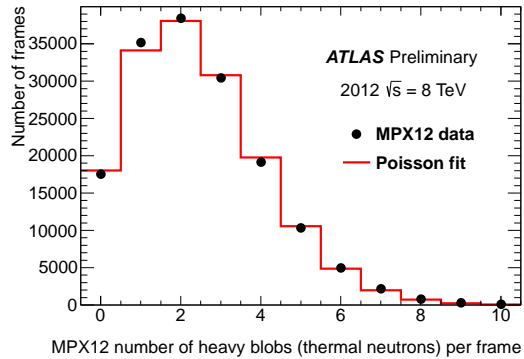}
\vspace*{-3mm}
\caption{Distribution of the number of heavy blobs per frame,
recorded with MPX12 below a $\mathrm{{}^{6}LiF}$ converter, 
during LHC collision periods in 2012.
The line shows a histogram with a fitted Poisson distribution.}
\label{fig:mpx12Neutrons}
\end{figure}

\begin{figure}[ht] 
\centering
\includegraphics[width=0.49\linewidth]{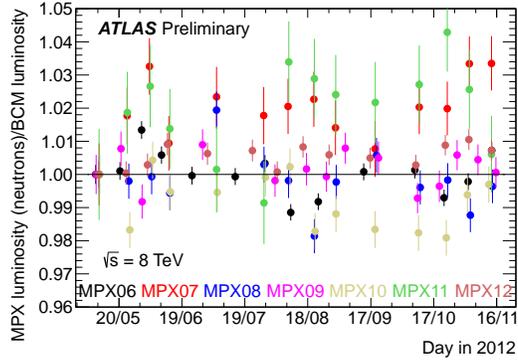}
\vspace*{-0.3cm}
\caption{Fractional deviation in the luminosity (thermal neutron counting mode)
reported by the MPX06 to MPX12 devices from the 
ATLAS reference luminosity (BCM) as a function of time in 2012. 
The ratio is shown for 14 time periods. 
For each device the time periods are chosen to have the same number of frames. 
The data are scaled such that the value of the first bin is unity.}
\label{fig:summaryNeutrons}
\end{figure}

Figure~\ref{fig:mpx12Neutrons} shows the distribution of the number of 
heavy blobs per frame below the $\mathrm{{}^{6}LiF}$ converter for MPX12.
The results from MPX luminosity measurements using thermal neutron counting 
are shown in Fig.~\ref{fig:summaryNeutrons} for the MPX06 to MPX12 devices.

There is no indication of  a significant slope 
from  the luminosity measured by the  heavy blob analysis 
(thermal  neutron counting)  in any of the MPX devices,
when normalized to the BCM measurement, 
although for some devices the statistical precision is not sufficient 
to test deviations at the percent level.
We recall that MPX06 indicates a 2\% slope for the luminosity 
from hit counting and the same device measured no slope for 
the luminosity from neutron counting. 
In the current analysis of the MPX data the difference between the 
luminosity monitoring using hits and  heavy blobs is  noted  
and further studies are required for an understanding of the underlying reason.

\section{Internal Consistency of MPX and Other ATLAS Relative-luminosity Measurements}
\begin{figure}[tp]
\centering
\begin{overpic}[width=0.49\linewidth]{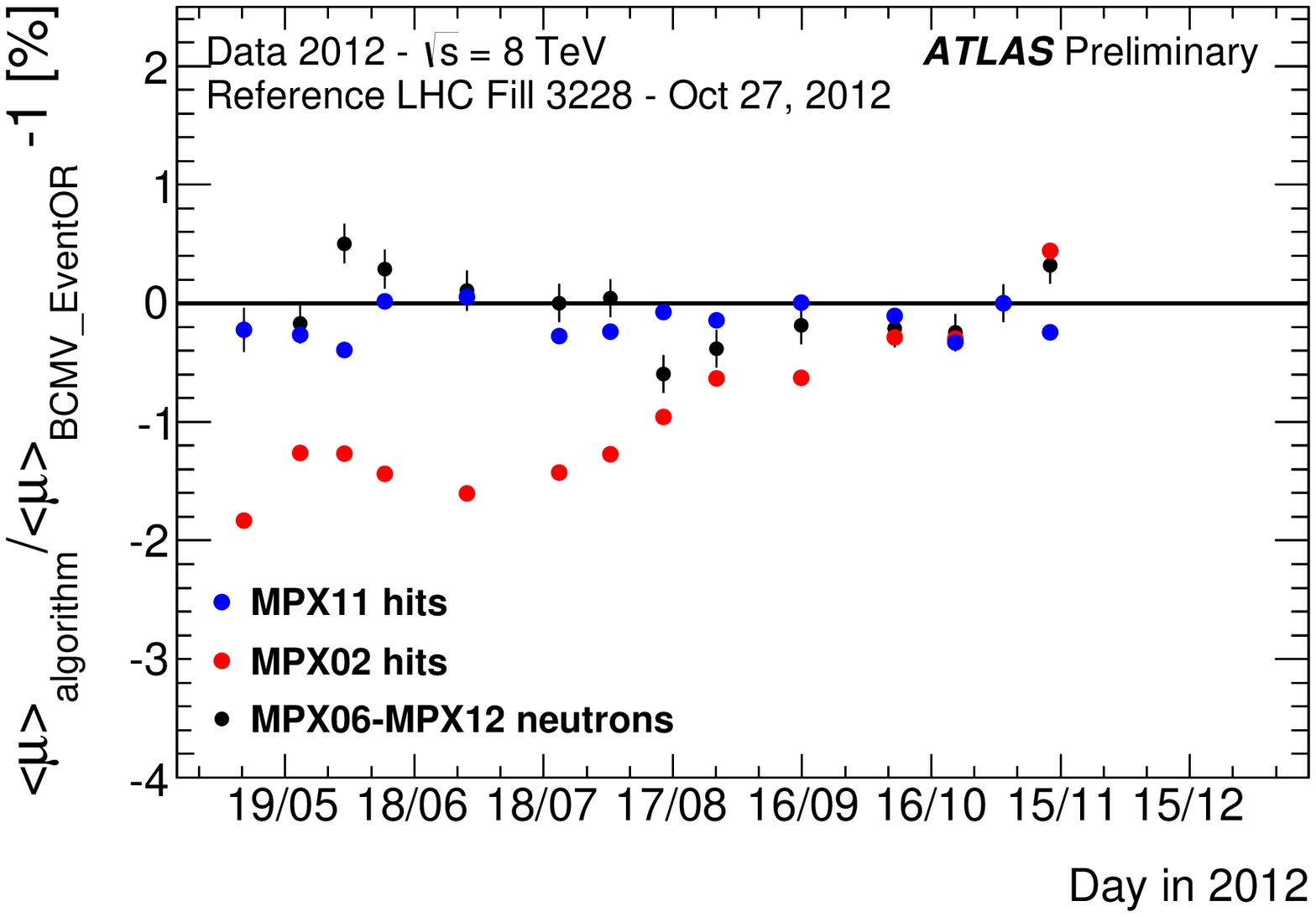}
	\put(86, 54.0){(a)}
\end{overpic}
\begin{overpic}[width=0.49\linewidth]{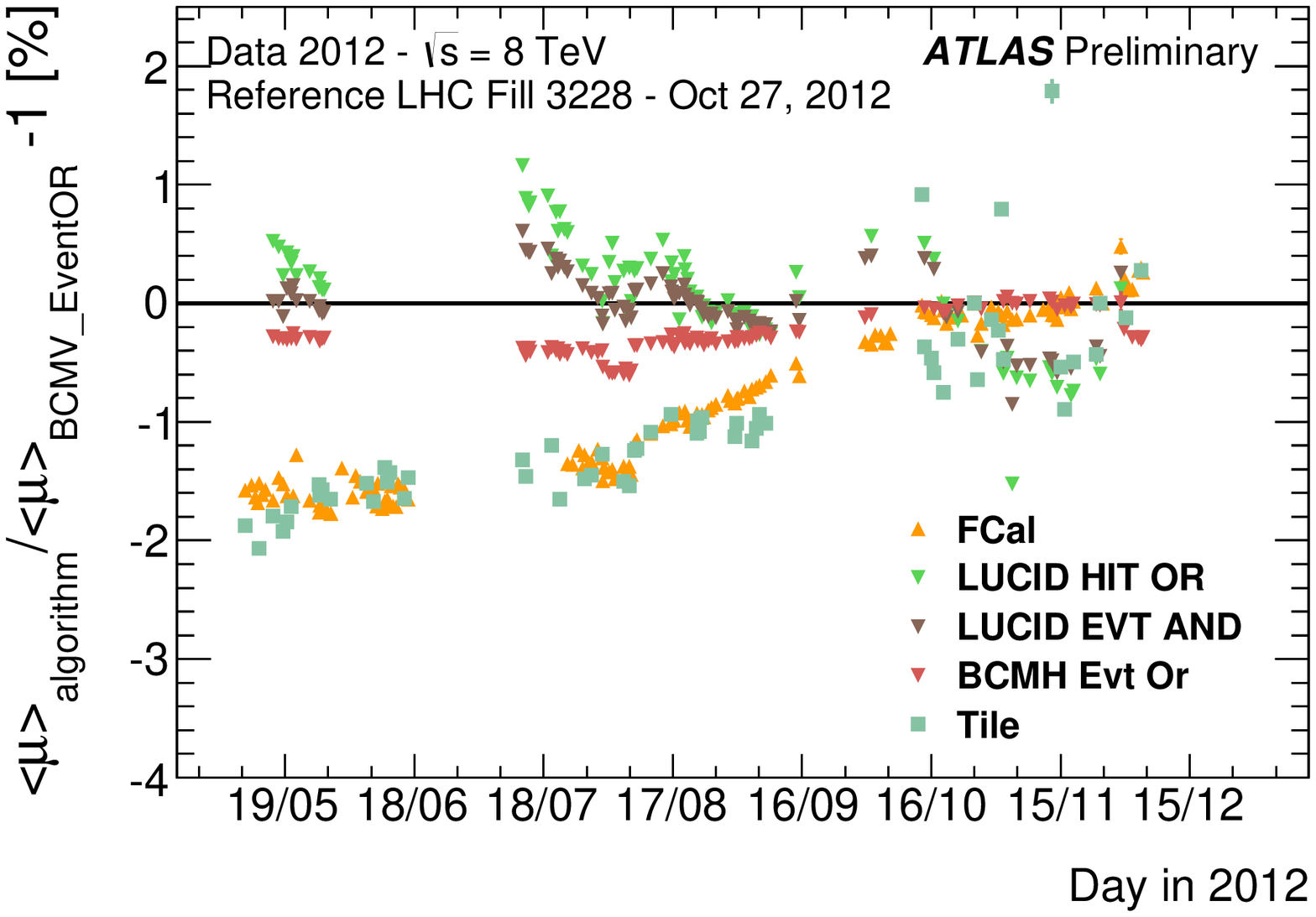}
	\put(86, 54.0){(b)}
\end{overpic}
\vspace*{-0.3cm}
\caption{Fractional deviation in the number of interactions per bunch crossing 
(averaged over all colliding bunch pairs), obtained using (a) MPX results 
and (b) other ATLAS luminometers, with respect to the $\rm BCMV\_EventOR$ 
value~\cite{improvedLumiDet:2013} and as a function of time in 2012. 
Each point shows the mean deviation for a single run compared to a 
reference run taken on October 27, 2012. Statistical uncertainties are shown 
per point, but in most cases are negligible.
The MPX hit luminosity from a device that gives a constant ratio (MPX11)
as well as that of a device with one of the largest increases (MPX02) are shown
together with the luminosity measured with the MPX from thermal neutron
counting averaged over MPX06 to MPX12.
}
\label{fig:fracDeviation}
\end{figure}

The relative long-term stability of the MPX luminosity measurements is compared to that 
of other \mbox{ATLAS} luminometers (BCM, LUCID, FCal and Tile)~\cite{improvedLumiDet:2013} 
in Fig.~\ref{fig:fracDeviation}a.
Both LUCID algorithms as well as the $\rm BCMH\_EventOR$ algorithm remain consistent, 
within about 1\%, with the reference algorithm $\rm BCMV\_EventOR$, 
as does the MPX-averaged neutron measurement (Fig.~\ref{fig:fracDeviation}b) 
and the MPX-hit luminosity reported by MPX11 
(see also Fig.~\ref{fig:summaryHits}c). 
In contrast, the calorimeter-based, bunch-averaged luminosity measurements display
a systematic drift of about 2\% with 
respect to the results of the reference algorithm over the 2012 running period,  
as do the MPX02 hit-luminosity measurements (Fig.~\ref{fig:summaryHits}a).
It is difficult to draw any clear conclusion since the full set of thirteen MPX
devices analyzed in this note have slopes as a function of time that span 
the range of slopes observed in the other ATLAS luminosity detectors.
Further analysis may identify specific features of these detectors that would 
explain the slope variations, but for now, an overall uncertainty of about 2\% 
appears to be required to describe the data.
Overall, the relative long-term stability of the MPX luminosity measurements 
is comparable to that of the other ATLAS luminometers.

\section{Van der Meer Scans}
\label{sec:vdM_scans}

Van der Meer (vdM) scans are used for absolute luminosity 
calibration at the LHC~\cite{improvedLumiDet:2013}.
In studies of \mbox{ATLAS} vdM data with MPX devices the focus is on the 
comparison  of the ratio of MPX  versus the BCM
luminosity measurement in order to quantify the relative stability of the 
BCM and MPX luminosity calibrations, 
both from one vdM scan session to the next (April, July and November 2012), 
and between the low-luminosity regime of vdM scans and the high-luminosity 
regime of routine physics running.
The beams are scanned transversely across each other in typically 25 scan steps. 
During each step, the beam orbits are left untouched (``quiescent beams") and the 
luminosity remains constant for approximately 29\,s. 
The beam separation is then incremented by several tens of microns 
(``non-quiescent" beams) over an interval of several seconds, 
during which the luminosity varies rapidly and the luminosity measurements are unreliable. 
As the MPX acquisition time is about 5\,s per frame, followed by a 6\,s readout period,  
typically two frames occur within each quiescent-beam scan step. 
Occasionally, the MPX devices need to reload their configuration files, 
in which case the dead time can be as long as 30\,s and therefore in some scan steps 
only one frame is recorded.

\begin{figure}[t]
\centering
\begin{overpic}[width=0.49\linewidth]{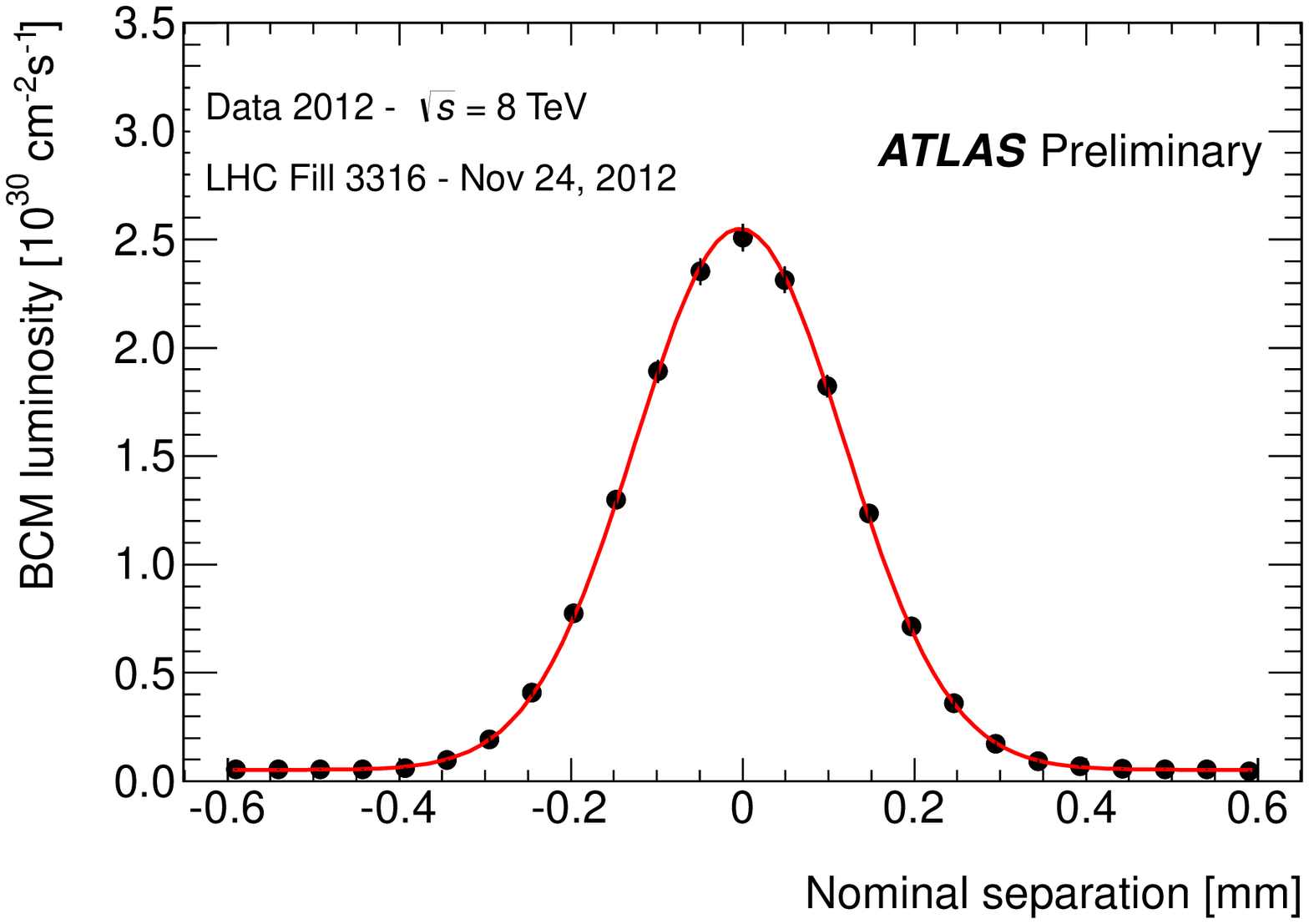}
	\put(86, 59.0){(a)}
\end{overpic}
\begin{overpic}[width=0.49\linewidth]{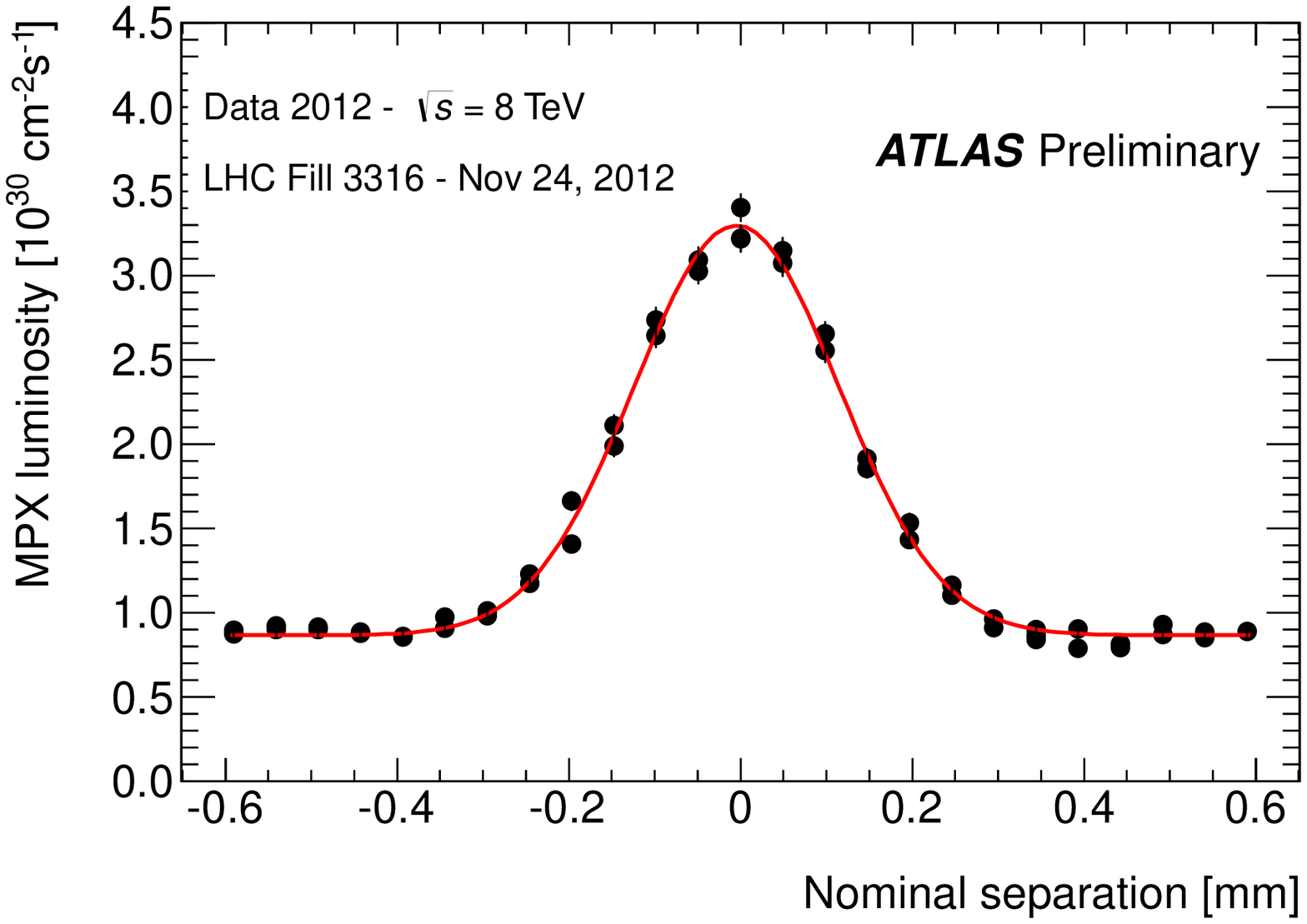}
	\put(86.0, 59.0){(b)}
\end{overpic}
\vspace*{-3mm}
\caption{(a) BCM luminosity during the last horizontal November 2012 vdM scan 
summed over all bunches as a function of nominal beam separation. 
Each data point shows the measured instantaneous luminosity 
before background subtraction and averaged over the entire duration of a scan step. 
Data from the scan steps during which the beam separation is varied 
(non-quiescent beams) are not shown. 
The fit function is the sum of the proper luminosity that is well represented 
in this scan by a single Gaussian 
and of a constant term that accounts for instrumental noise and single-beam background.
(b) MPX01 hit luminosity during the last horizontal November 2012 vdM scan 
summed over all bunches as a function of nominal beam separation. 
Each data point shows the measured instantaneous luminosity 
before background subtraction and averaged over one sampling interval 
of the MPX01 device. 
The MPX samplings that partially or totally overlap with non-quiescent 
scan steps (varying beam separation) are not shown. 
The fit function is the sum of the proper luminosity that is represented 
in this scan by a single Gaussian 
and of a constant term that accounts for instrumental 
noise and single-beam background.
}
\label{fig:mpx01HitLumi2}
\end{figure}

The beam-separation dependence of the measured luminosity is 
reasonably well represented by the sum of a single Gaussian plus a 
constant for both the BCM data 
(Fig.~\ref{fig:mpx01HitLumi2}a). 
and the MPX data
(Fig.~\ref{fig:mpx01HitLumi2}b). 
The data show that the sensitivity of the MPX01 device is sufficient to measure the 
luminosity down to about $5 \times 10^{29} {\rm \,cm}^{-2} {\rm s}^{-1}$, 
four orders of magnitude smaller than the luminosity typical of routine physics running.
The sensitivity of the hit method, therefore, potentially allows the MPX 
to be used to cross-check the calibration transfer of other 
luminometers (BCM, LUCID, Tile) from van der Meer scans to physics 
conditions.

\begin{figure}[bhp]
\centering
\includegraphics[width=0.49\linewidth]{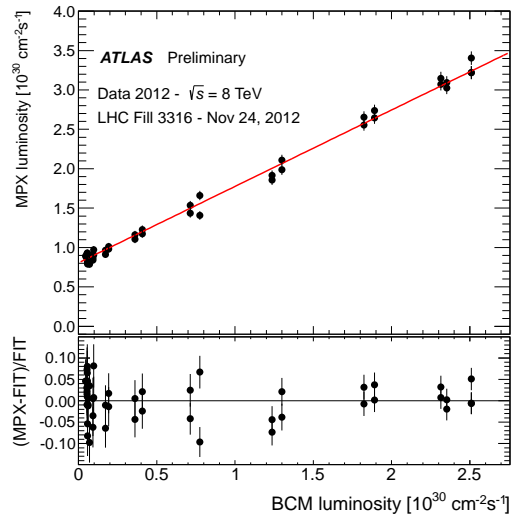}
\vspace*{-3mm}
\caption{MPX01 hit luminosity vs. BCM luminosity for the last 
horizontal November 2012 vdM scan. 
The line is a linear fit to the data. 
The non-zero intercept corresponds to the difference of the MPX and BCM 
noise and background contributions. 
The bottom frame shows the fractional residual of the linear fit.
}
\label{fig:bcmMpx01Lumi}

\end{figure}

Figure~\ref{fig:bcmMpx01Lumi} demonstrates that, on average, 
the MPX response is a linear function of the ATLAS reference (BCM) luminosity down to very 
low luminosities ($\rm \approx 10^{29}\,cm^{-2}s^{-1}$), albeit with fluctuations of up 
to 10\% (at constant luminosity) that cannot be explained by statistical 
arguments. 
It has been established that the fluctuations arise from the MPX measurements 
rather than the BCM measurements by comparing scan step by scan step and by comparing
the luminosity values reported by multiple BCM and LUCID algorithms 
which are consistent within a fraction of a percent.
Similar fluctuations of the MPX response, also of apparently non-statistical origin, 
have been observed at much higher luminosity during routine physics running. 
These issues remain under study.

\clearpage
\section{Conclusion}
\label{sec:conclusion}

The network of  MPX  devices installed in ATLAS has 
successfully taken data from 2008 to 2013  and proven to be 
sufficiently radiation  hard for the high-luminosity data taken 
in 2012 at $\sqrt{s}=8$~TeV proton-proton collisions.  
The study presented here focuses on 
luminosity measurements during proton-proton collisions from 
April  to December 2012.  The  study has demonstrated that the 
MPX  network has an internal consistency of about 2\%  
using different detectors and techniques 
for hit counting and heavy blob (thermal neutron) counting as 
measures of luminosity. 
This number is comparable to the present preliminary discrepancy 
found in the long-term comparisons between other luminosity detectors in ATLAS
in the same time period.
In addition, the MPX network has been used to study in detail 
the three van der Meer scan periods performed in 2012, 
in a regime where the luminosity is four orders of magnitude lower 
than during routine physics data-taking.
Although not specifically designed for 
luminosity measurements in ATLAS, the  MPX  network  gives
reliable supplementary information for the overall   ATLAS 
luminosity determination  over a wide dynamic range 
(luminosities from about 
$\rm 5 \times 10^{29}\,cm^{-2}s^{-1}$ to $\rm 7 \times 10^{33}\,cm^{-2}s^{-1}$). 
The present study  represents  a proof of principle and will  be extended to 
address systematic uncertainties  with the aim to increase the 
precision of the MPX luminosity monitoring using the
accumulated high-statistics data set.

\acknowledgments

We would like to thank the organizer of the 
13th Topical Seminar on Innovative Particle and Radiation Detectors 
(IPRD13) for their kind invitation.


\bibliographystyle{atlasnote}
\bibliography{biblio}

\end{document}